**Demand-side decarbonisation at scale via MaaS-integrated carbon incentives**

Bing Liu[1], Yuan Liao[2,3], Sonia Yeh[2], Oded Cats[4], Kristian S. Nielsen[5], Zhenning Dong[6], Yong Wang[7], Yanli Liu[8], Zirui Ni[1], Xiaolei Ma[1,9*]

*[1] School of Transportation Science and Engineering, Beihang University, Beijing 100191, China*
*[2] Department of Space, Earth and Environment, Chalmers University of Technology, 41296 Göteborg, Sweden*
*[3] Department of Applied Mathematics and Computer Science, Technical University of Denmark, Lyngby, Denmark*
*[4] Department of Transport and Planning, Delft University of Technology, 2628 CD Delft, Nederland*
*[5] Department of Management, Copenhagen Business School, Frederiksberg, Denmark*
*[6] AutoNavi Software Company, Ltd, Beijing 100020, China*
*[7] School of Economics and Management, Chongqing Jiaotong University, Chongqing 400074, China*
*[8] Key Laboratory of Smart Grid of Ministry of Education, Tianjin University, Tianjin 300072, China*
*[9] Key Laboratory of Intelligent Transportation Technology and System, Ministry of Education, Beijing 100191, China*

* Corresponding author. Email: xiaolei@buaa.edu.cn

## SCIENCE FOR SOCIETY

Cities need to cut transport emissions without restricting access to jobs and services. Digital mobility platforms can reward low-carbon travel, but incentives cannot compensate for poor transport options. We evaluate Beijing's CIP-MaaS program, which links recorded low-carbon trips, personal carbon accounts, platform rewards, and carbon-market revenue. Enrollment was associated with a higher share of public transport and cycling trips, with larger responses in areas offering better transit access. For decision makers, the findings highlight the importance of credible carbon accounting, transparent use of carbon revenue, and coordination between incentives and transport investment. Programs should also monitor unequal access, as residents of poorly served areas may have fewer opportunities to respond or benefit. Digital incentives should therefore complement, rather than replace, reliable public transport and cycling infrastructure.

## Highlights

- MaaS-integrated carbon incentives are linked to city-scale mobility shifts
- 3.9 million participants increased low-carbon travel after enrollment
- Carbon-accounting scenarios imply ~94,000 tonnes $CO_2$ annually
- Larger responses concentrate where transit access is stronger

## SUMMARY

Digital carbon incentives are increasingly used to promote low-carbon travel, but city-scale evidence on their behavioral and carbon-accounting implications remains limited. We evaluated a carbon-incentive program on a Beijing Mobility-as-a-Service platform, using 4.82 billion trips from 2.96 million users over 13 months, with a matched panel of enrolled and non-enrolled users. Enrollment was associated with a 20.3-percentage-point increase in monthly low-carbon travel share, with pre-enrollment trends near zero across event-study tests. A pre-enrollment-trained random-forest accounting scenario implied a 1.8% citywide decline in gasoline-car trips and 94,353 tonnes $CO_2$ annually, equivalent to 5.7% of certified reductions traded in Beijing's carbon market in 2023; this value is model-dependent rather than a directly observed or causally identified program effect. Larger program-associated responses were concentrated in areas with greater transit access. These results show that carbon-market-financed digital incentives can support measurable low-carbon travel responses at city scale.

## Introduction

Meeting global carbon reduction targets requires rapid decarbonization of the transport sector, which accounts for nearly a quarter of global $CO_2$ emissions[1]. In cities, transport emissions are closely linked to local air pollution, noise, congestion, and inefficient land use, creating environmental and health burdens that often fall unevenly across population groups[2,3]. Reducing urban transport emissions therefore requires not only cleaner vehicle technologies, but also changes in everyday mobility practices that sustain high-carbon travel.

Urban transport decarbonization depends on three interrelated dimensions: technology, infrastructure, and behavior[4,5]. Electrification can reduce emissions from motorized travel, and investments in public transport, cycling networks, and transit-oriented development can expand the set of feasible low-carbon alternatives. However, private-car dependence remains deeply embedded in many urban systems, limiting the extent to which infrastructure and technology alone can reduce emissions. Demand-side interventions that influence mode choice, trip planning, and repeated travel behavior may therefore complement infrastructure investment and vehicle electrification, especially when low-carbon alternatives are already accessible.

A wide range of policy instruments has been used to influence travel behavior, including congestion charges, parking restrictions, fuel taxes, and subsidies for public and active transport[6,7]. Digital mobility platforms and Mobility-as-a-Service (MaaS) systems offer a newer operational channel for such interventions. By integrating route planning, travel information, booking, and payment across modes, MaaS platforms can also deliver incentives, feedback, and behaviorally salient information through a common user interface[8,9]. Yet the effectiveness of digitally delivered mobility incentives remains uncertain. Existing evidence is often based on small-scale pilots, short observation windows, or self-reported travel behavior, which limits the ability to measure changes in actual travel patterns and associated emissions at city scale[10–14].

Prior MaaS and mobility-incentive studies have reported mixed findings. Some pilots suggest reductions in private-car use, whereas others show limited or context-dependent behavioral responses[15–17]. Survey-based studies have also identified heterogeneous interest in MaaS across age, education, income, and environmental attitudes, but such evidence is vulnerable to social desirability bias, recall error, and the gap between stated preferences and observed behavior[18–24]. More broadly, many studies lack high-frequency longitudinal trip records that capture travel frequency, timing, mode choice, and multi-stage trip chains, all of which are needed to estimate transport-related carbon emissions with sufficient behavioral detail[25–30]. As a result, there remains limited real-world evidence on whether digital mobility incentives are associated with measurable low-carbon travel responses, how these responses evolve during active implementation, and under what urban conditions they are stronger.

This study addresses this gap through a large-scale real-world evaluation of Beijing's Carbon Incentive Program integrated within a MaaS platform (CIP-MaaS). Implemented through the Gaode digital map app, the program links eligible low-carbon travel activity to verified carbon accounting and platform-based rewards[31]. Between December 2019 and December 2023, the program registry recorded 3,983,027 cumulative registrations. The behavioral dataset used in this study covers December 2022 to December 2023 and contains 4.82 billion valid multimodal trip records from 2,958,841 observed platform users. For the primary behavioral analysis, we constructed a matched longitudinal panel of enrolled participants and non-enrolled platform users. This combination of program-registration, mobility, and matched-panel data provides a large-scale setting for examining a citywide digital carbon-incentive program using observed rather than self-reported travel behavior.

Using matched longitudinal data and a difference-in-differences design, we estimate program-associated changes in mode choice and $CO_2$ emissions, examine how low-carbon travel responses evolved while incentives remained active, and identify individual and urban characteristics associated with larger estimated responses. We find that enrollment in CIP-MaaS was associated with a substantial increase in the monthly low-carbon travel ratio. A secondary random-forest accounting analysis generated model-implied differences in gasoline-car travel and scenario-based carbon-accounting values during the observation period. The response attenuated over time but remained elevated while incentives were active, suggesting sustained engagement rather than evidence of post-incentive habit formation. Spatial analyses further showed that larger program-associated responses were concentrated in areas with greater public-transport accessibility, highlighting the need to coordinate digital incentives with physical low-carbon travel alternatives. We therefore interpret the findings as quasi-experimental, program-associated evidence on an active digital carbon-incentive program, while recognizing that voluntary enrollment and the absence of post-incentive observations limit strong causal and persistence claims.

## Results

### Overview of the CIP-MaaS program and the data

Using high-frequency mobility records from millions of individuals, we evaluate travel-behavior changes, estimated $CO_2$ reductions, and spatial response patterns associated with enrollment in the CIP-MaaS program in Beijing, China. The program links individual low-carbon travel activity to verified carbon accounting through a citywide MaaS platform operated via the Gaode (AutoNavi) digital map app in collaboration with public transport agencies[32–34]. Low-carbon trips, including public transport and cycling, are recorded through the platform and public transport transaction systems, credited to individual carbon accounts, and connected to platform-based rewards. At the program level, verified reductions are aggregated and monetized through the local carbon market, and the resulting revenue supports the reward system (Fig. 1, Note S1). Synchronization between the public transport corporation and the Gaode backend improves consistency in low-carbon trip accounting and reduces the risk of duplicate or missing records.

Our analysis integrates 13 months of trip-level data from December 2022 to December 2023. The final trip dataset comprised 4.82 billion valid records from 2.96 million observed platform users, while the CIP-MaaS registry contained 3.98 million cumulative registrations between December 2019 and December 2023. Aggregate comparisons indicated that the Gaode dataset captured a substantial share of estimated citywide travel and had a broadly similar, although not identical, modal distribution to the citywide benchmark (Tables S1 and S2). Separately, we compared public-transit-related trip frequencies from the Gaode backend with an independent transaction dataset from the Beijing Public Transport Corporation. Equivalence tests using a pre-specified 10% margin showed closely aligned public-transport travel patterns across the two recording systems for the examined user groups (Fig. S2). These comparisons support the use of the platform records while not implying that platform users or CIP-MaaS registrants were fully representative of all Beijing residents.

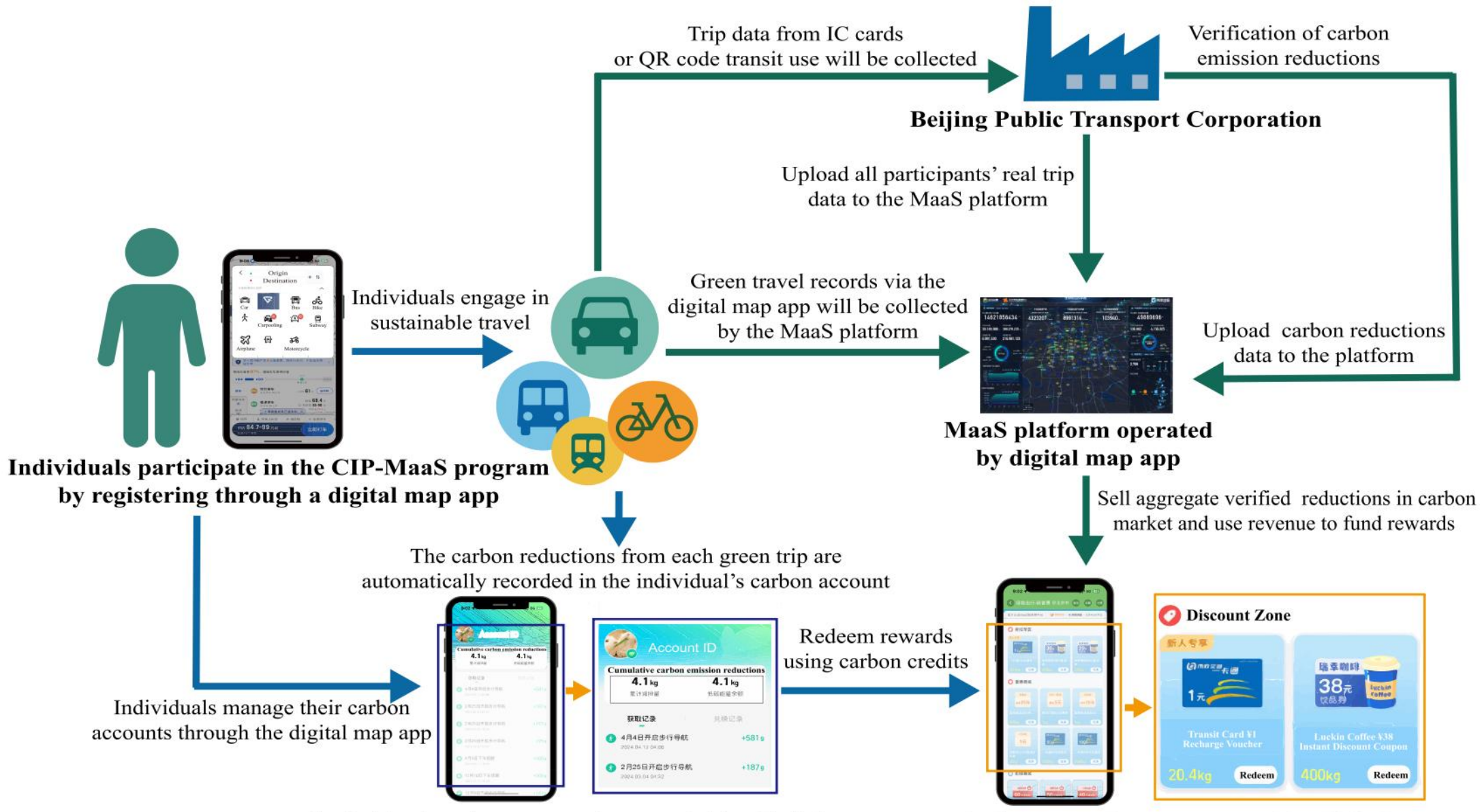

**Fig. 1 Operational mechanism of the CIP-MaaS program through the Gaode app.** Individuals enroll in the CIP-MaaS program through the Gaode app. Low-carbon trips, including public transport and cycling, are recorded and verified through the MaaS platform and public transport transaction systems. Verified emission reductions are credited to individual carbon accounts and can be redeemed for rewards through the platform. At the program level, verified reductions are aggregated and monetized through the carbon market, and the resulting revenue is used to support the reward system.

We also used financial records to characterize the funding basis and treatment magnitude of the incentive mechanism. PCER transactions dated within the behavioral analysis window certified 97,635 tonnes $CO_2$ and generated 5.37 million CNY at 55 CNY per tonne $CO_2$ (Note 7). Reward-redemption exposure was summarized by event month relative to individual enrollment. Across event months 0–8, redeemed platform rewards totaled 5.02 million CNY, averaging 7.94 CNY per active participant-month. The redeemed value was equivalent to 93.47% of the certified PCER revenue included in the analysis (Tables S29 and S30). Because certification, carbon-market transactions, and reward redemption did not necessarily occur contemporaneously, this ratio is interpreted as a descriptive measure of financial scale rather than an exact cash-flow balance.

**The MaaS incentive program increased low-carbon travel**

To estimate program-associated changes in travel behavior, we constructed a matched longitudinal panel containing 1,897,129 unique platform users: 632,376 CIP-MaaS participants and 1,264,753 non-enrolled platform users. Propensity-score matching was used to improve comparability between the two groups, and a difference-in-differences model was applied to the resulting 17,841,159 user-month observations (Note S4 and Figs. S3 – S5). Enrollment was associated with a 20.3 percentage-point increase in the monthly low-carbon travel ratio relative to the matched comparison group (Table S9).

Event-study estimates for the available pre-enrollment months were close to zero (Fig. 2A and Table S3). This pattern is consistent with the absence of large differential pre-enrollment trends but does not establish the parallel-trends assumption. After enrollment, participants' low-carbon travel ratio was higher than that of matched non-participants. The estimated difference was 20.4 percentage points in the enrollment month and remained elevated at 12.8 percentage points eight months after enrollment (Fig. 2A). Placebo analyses assigning hypothetical enrollment dates one or two months earlier produced no statistically significant estimates (Tables S4–S5). Together, these diagnostics are consistent with a program-associated increase in the observed low-carbon travel ratio, while not ruling out all sources of non-random selection.

As an additional robustness analysis, we estimated a staggered difference-in-differences specification using not-yet-treated future participants as the comparison group. The model-based average post-enrollment ATT was 19.5 percentage points (95% CI: 19.17 – 19.89), and the joint pre-trend test did not reject the null of no differential pre-enrollment trends ($P = 0.125$; Fig. S6 and Tables S6 – S7). Oster coefficient-stability analysis indicated that the DID estimate was highly stable to adjustment for observed covariates. Under the recommended bound Rmax = 1.3 × $\tilde{R}$, selection on unobserved variables would need to be approximately 130.7 times as strong as selection on observed covariates to explain away the estimate, while the bias-adjusted estimate remained positive at 20.05 percentage points (Table S8).

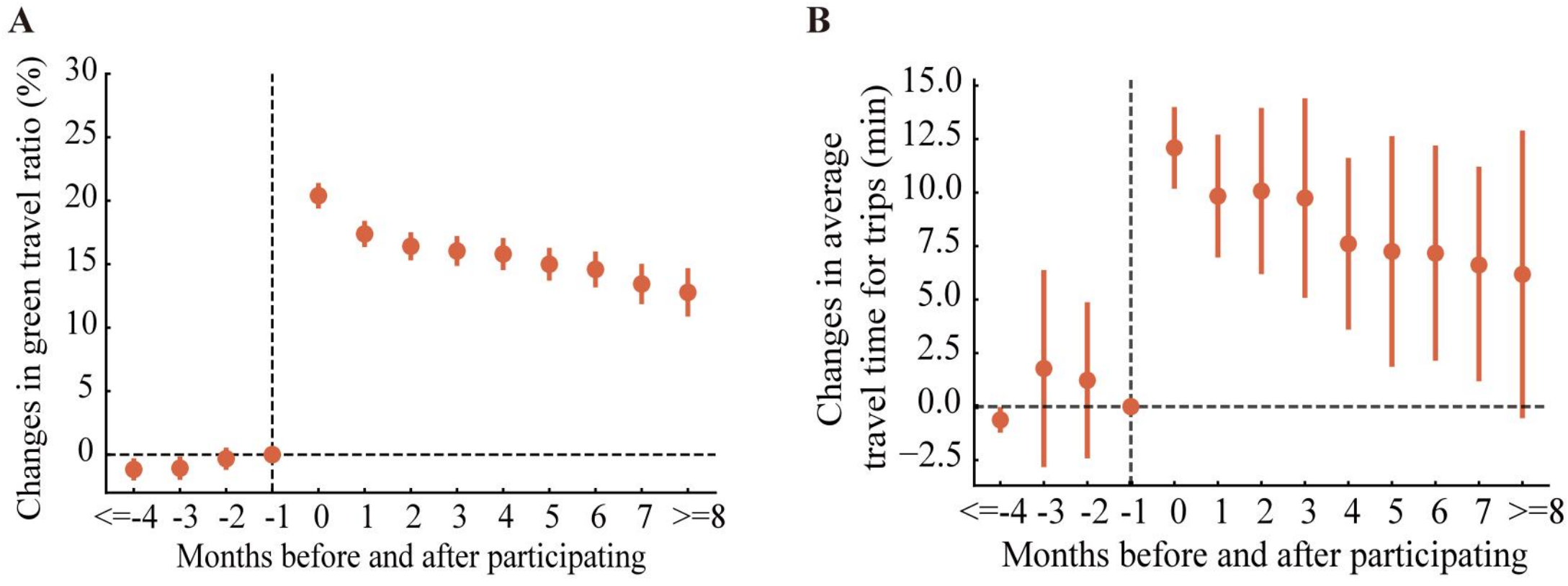

**Fig. 2 Estimated changes in travel patterns associated with CIP-MaaS enrollment.** (A) Estimated changes in the monthly low-carbon travel ratio, with the month before program enrollment used as the reference period. (B) Estimated changes in average trip duration.

Beyond the low-carbon travel ratio, we also examined changes in participants' travel time and distance (Fig. 2B and Tables S16–S19). Average trip duration increased by 7.4 minutes ($P = 0.003$) and trip distance by 2.8 km ($P = 0.010$) immediately after enrollment, suggesting that some low-carbon travel responses involved longer observed journeys, consistent with greater use of public transport and cycling. These travel-time and distance effects attenuated gradually over time. Nevertheless, the monthly low-carbon travel ratio remained elevated throughout the observation window, indicating continued engagement with low-carbon travel while the incentive program was active. These results should therefore be interpreted as changes in the composition and characteristics of observed trips, rather than as a full decomposition of mode substitution, trip frequency, and destination choice.

**Heterogeneity in estimated responses**

We next examined how estimated responses varied across socio-demographic and behavioral groups (Fig. 3). Larger estimated responses were observed among women, young adults, users in the higher mobile-phone-value proxy group, and individuals with lower monthly trip frequencies (Tables S10–S15). Female participants increased their share of low-carbon trips by 21.3 percentage points, compared with 19.6 percentage points for males (Fig. 3A). Participants aged 19–24 years showed the largest estimated response, with a 30.7 percentage-point increase in low-carbon travel (Fig. 3B). Estimated responses also differed across phone-value proxy groups, with users in the higher phone-value proxy group increasing their low-carbon travel ratio by 23.5 percentage points. Participants with lower monthly trip frequencies and those with average trip durations of 45–60 minutes also showed relatively large estimated responses (26.0 and 21.0 percentage points, respectively; Fig. 3D–E).

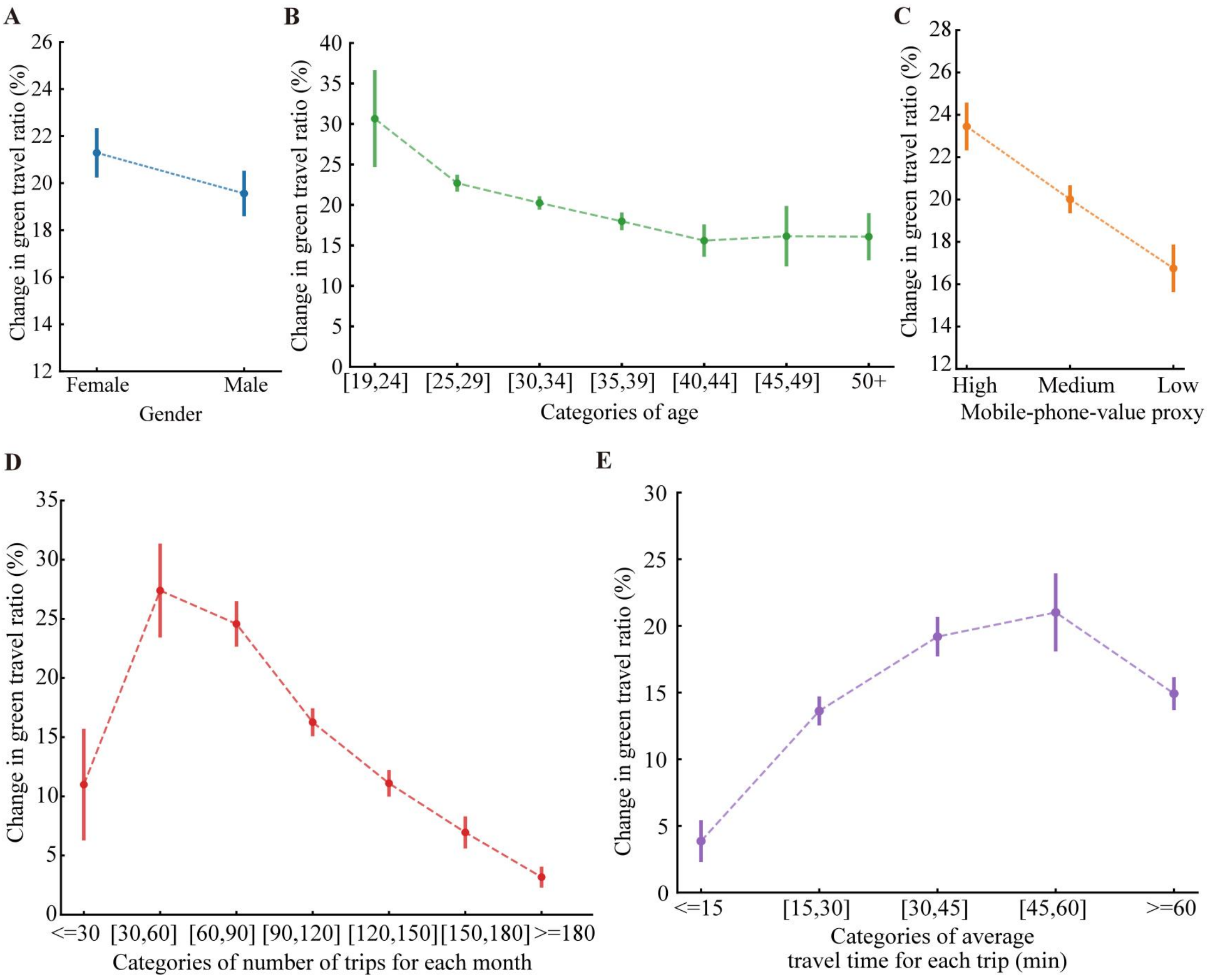

**Fig. 3 Heterogeneous estimated responses across population segments.** Subgroup-specific estimated changes in the monthly low-carbon travel ratio by (A) gender, (B) age, (C) mobile-phone-value proxy, (D) monthly trip frequency, and (E) average travel time per trip. Values on the y-axes are reported in percentage points. Detailed estimates are presented in Tables S10–S15. Proxy categories were derived from estimated mobile-phone value and do not represent reported income.

**Scenario-based carbon accounting of model-implied mode shifts**

Carbon-account records report credited reductions from eligible low-carbon trips but cannot determine which trips occurred because of CIP-MaaS. We therefore treat the matched difference-in-differences estimate of the monthly low-carbon travel ratio as the primary evidence of program-associated behavioral change. As a secondary accounting exercise, we trained a random-forest classifier on participants' pre-enrollment trips and applied it to post-enrollment trips to generate a model-classified reference mode under comparable recorded trip characteristics (Table S20 and Figs. S7–S8). The predictors included departure time, origin, destination, realized trip duration, realized trip distance, calendar date, and workday status. Because realized duration and distance may depend on the observed travel mode, the classified reference mode should not be interpreted as an independently identified untreated counterfactual. The resulting calculations are therefore reported as model-dependent scenario estimates rather than causal estimates of program-attributable emission reductions.

Comparison of observed post-enrollment modes with the modes classified by the pre-enrollment-trained random-forest model produced model-implied differences of 4.2%, 27.6%, 33.5%, and −14.1% for bicycle, bus, subway, and gasoline-car trips, respectively (Fig. 4A–H). Because participants accounted for an estimated 12.6% of daily trips in Beijing, scaling these differences to the citywide level produced corresponding scenario values of 0.5%, 3.5%, 4.2%, and −1.8%. Thus, under the specified accounting scenario, the observed post-enrollment trip patterns corresponded to a model-implied 1.8% citywide decline in gasoline-car trips (Table S1). These values represent differences between observed and model-classified travel modes rather than directly observed citywide changes or causal program effects.

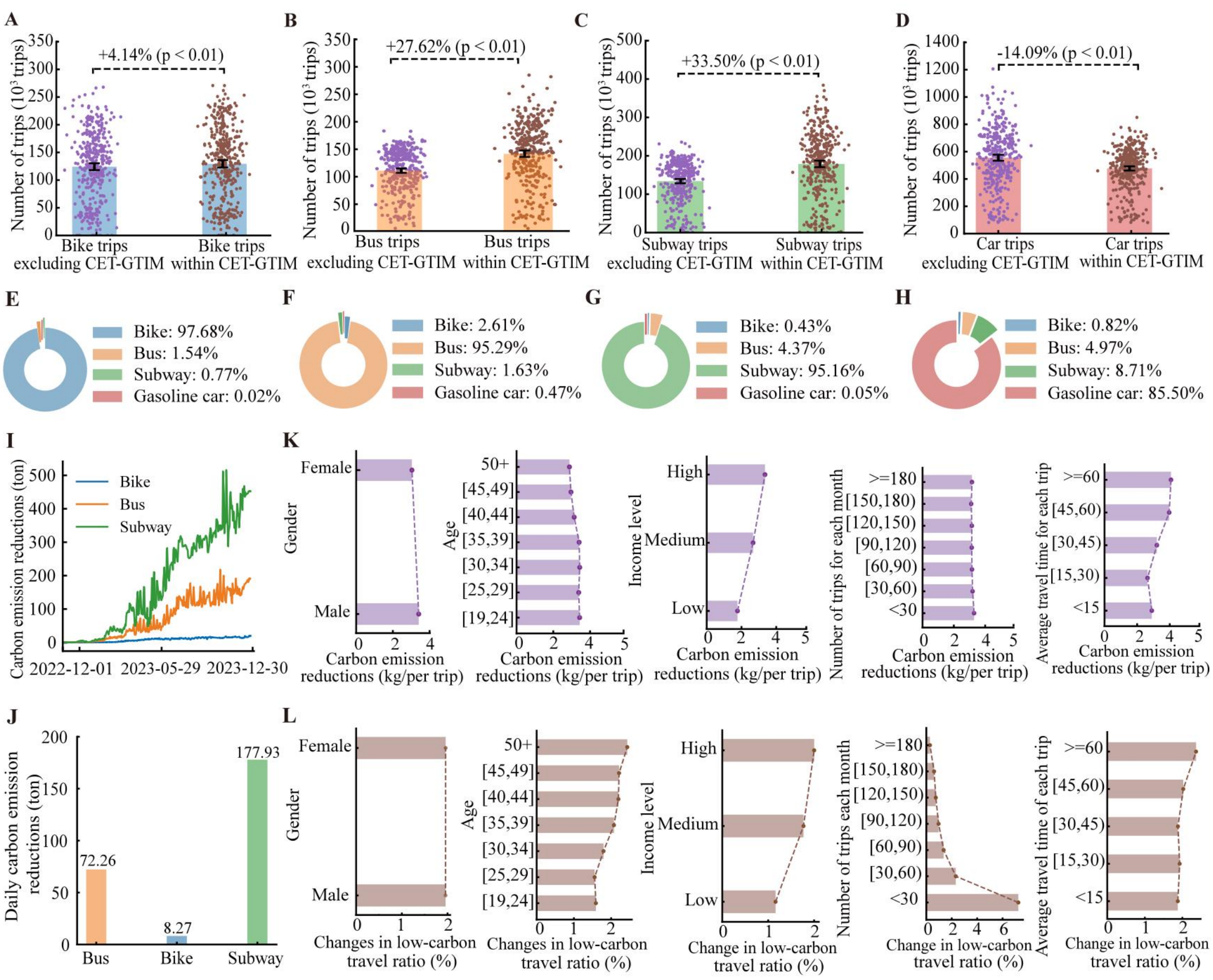

**Fig. 4 Model-based comparison of observed and classified travel modes and associated**

**scenario-based carbon accounting.** Panels (A)–(H) compare observed post-enrollment travel modes with reference modes classified by a random-forest model trained on pre-enrollment trips. Panels (I) and (J) report carbon-accounting values for trips assigned a gasoline-car reference mode but observed as cycling, bus, or subway. Panels (K) and (L) summarize corresponding descriptive values across participant groups. The model-implied mode differences corresponded to a 1.8% citywide decline in gasoline-car trips and an annualized gross carbon-accounting value of 94,353 tonnes $CO_2$, equivalent in scale to 5.7% of certified reductions traded in Beijing in 2023. Because the classified reference modes are predicted rather than observed untreated outcomes, these panels represent model-dependent accounting scenarios rather than direct causal estimates.

Under the pre-enrollment-trained random-forest accounting scenario, 14.5% of trips assigned a gasoline-car reference mode were observed as public transport or cycling, whereas the reverse model-implied shift was below 0.5% (Fig. 4E–H). Trips classified as gasoline-car travel but observed as subway, bus, or cycling yielded gross accounting values of 177.9, 72.3, and 8.3 tonnes $CO_2$ $day^{-1}$, respectively (Fig. 4I–J). Summed across the three modes and annualized, the procedure produced a carbon-accounting value of 94,353 tonnes $CO_2$, equivalent in scale to 5.7% of the certified reductions traded in Beijing's carbon market in 2023[35]. These quantities are conditional on the observed trips, model predictors, classification rule, and emission factors. They should therefore be interpreted as model-dependent scenario values rather than directly observed or causally identified program effects.

We further summarized scenario-based carbon-accounting values across participant groups (Figs. 4K–L and Table S21). Under the specified accounting procedure, the mean value per trip assigned a gasoline-car reference mode and observed as low-carbon travel was 3.41 kg $CO_2$ for men and 3.04 kg $CO_2$ for women; values also varied across age, income-proxy, trip-frequency, and travel-duration groups. These differences reflect the observed post-enrollment trip characteristics and the model-classified reference modes used in the accounting procedure. They are descriptive and should not be interpreted as causal differences in program responsiveness, realized mitigation effectiveness, or the expected benefits of targeting specific population groups.

**Spatial distribution and urban correlates of program-associated responses**

We examined the spatial distribution of program-associated response indicators across 500 m × 500 m urban zones. A graph convolutional network was used to learn zone embeddings from multimodal mobility flows, average trip times, travel distances, and zone-level response indicators. Hierarchical clustering of the learned embeddings yielded four groups, which we labeled high-response (HI), moderate-response (MI), low-response (LI), and negligible-response (NI) zones (Fig. 5A–C, Fig. S9, and Table S22). These labels were assigned partly according to gasoline-car travel reduction, estimated $CO_2$ reduction, and model-implied mode-shift indicators. Consequently, differences in these outcomes across HI, MI, LI, and NI zones are descriptive features of the cluster construction and do not provide independent evidence that infrastructure moderated the program response.

By construction, the four spatial groups displayed different model-implied response levels. The gasoline-car travel difference was 15.1 percentage points in HI zones, compared with 7.8, 4.2, and 0.01 percentage points in MI, LI, and NI zones, respectively. The corresponding average daily scenario-based $CO_2$ accounting values were 57.5, 18.7, 10.1, and near-zero kg $CO_2$ per zone, respectively (Fig. 5D–F and Figs. S10–S11). These contrasts summarize the outcome-based cluster definitions and should not be interpreted as independent tests of accessibility effects.

Separately, descriptive characteristics showed that HI zones were closer to the urban center and had greater average subway-network density than the other spatial groups (Table S28). Taken together, the spatial distributions show that larger program-associated responses were concentrated in areas with greater transit accessibility.

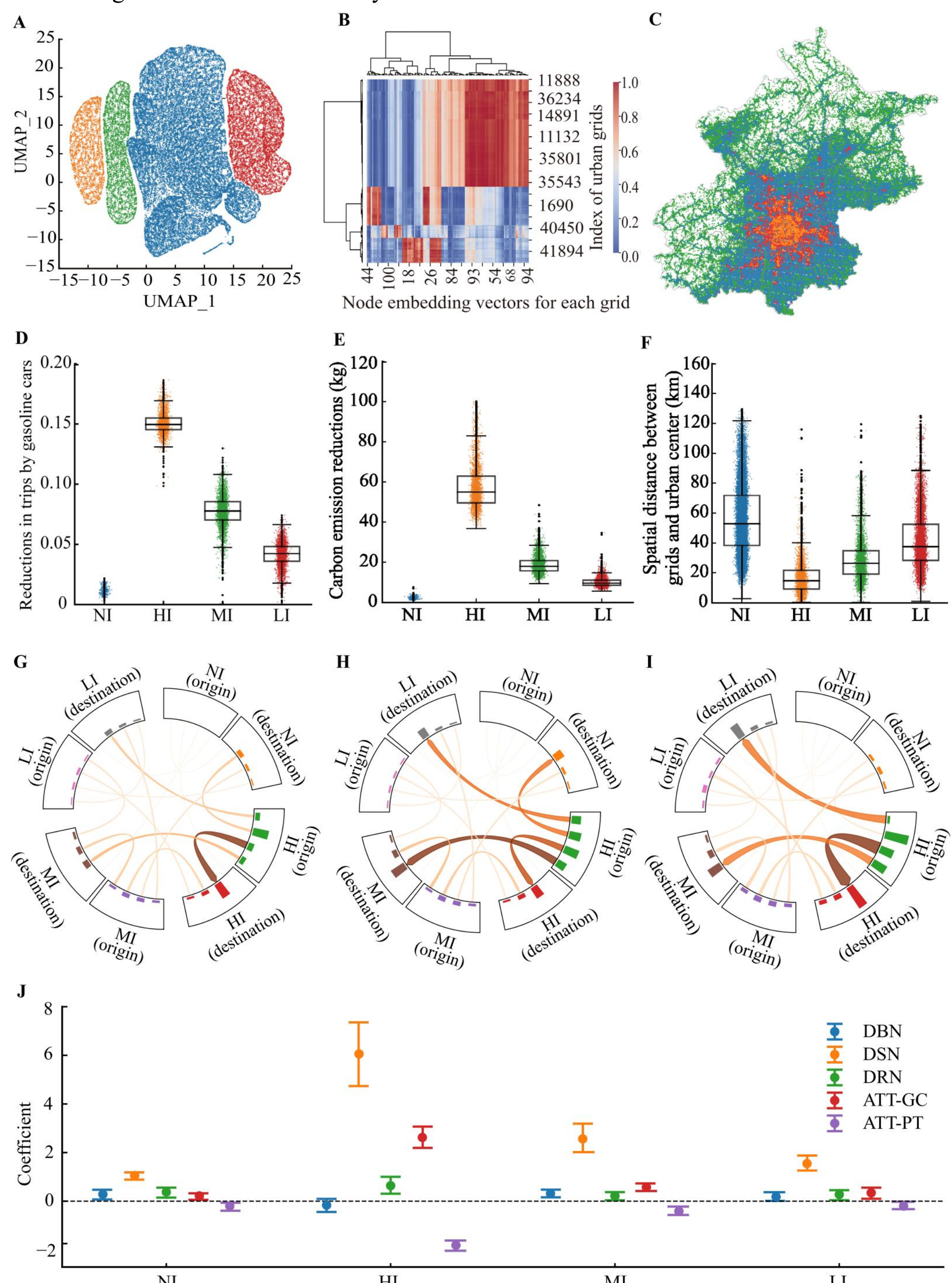

**Fig. 5 Spatial distribution and urban correlates of program-associated response indicators.** (A) Uniform Manifold Approximation and Projection visualization of urban-zone embeddings. (B) Hierarchical clustering of urban zones. (C) Spatial distribution of the resulting clusters. (D)–(F) Model-implied gasoline-car travel differences, scenario-based $CO_2$ accounting values, and distance from the urban center across cluster groups. (G)–(I) Model-implied shifts from gasoline cars to bicycles, buses, and subways. (J) Within-cluster associations between response indicators

and urban characteristics. The HI, MI, LI, and NI labels were assigned partly according to the response indicators shown in the figure. Because response indicators informed the cluster labels, between-cluster differences are descriptive; within-cluster regressions summarize correlations with urban characteristics.

Model-implied mode-shift patterns also differed across the outcome-defined spatial groups (Fig. 5G–I). Gasoline-car-to-bicycle and gasoline-car-to-subway shifts were more concentrated in HI zones, while gasoline-car-to-bus shifts were more common in HI and MI zones. Because mode-shift indicators contributed to the characterization of the clusters, these between-group contrasts are descriptive rather than independent evidence of infrastructure moderation. We subsequently fitted separate ordinary least-squares regressions within each cluster, relating zone-level gasoline-car travel and scenario-based $CO_2$ response indicators to road, bus, and subway network densities and average travel times. Subway-network density was positively associated with both response indicators across the cluster-specific models (Fig. 5J and Tables S23–S27). These regressions provide correlational support that stronger responses were co-located with greater subway-network density, complementing the descriptive spatial distributions.

The demographic and travel profiles of the spatial groups provide additional descriptive context (Table S28). Trips associated with HI zones had shorter average travel times and involved higher shares of women, users aged 19–34, and users in the middle phone-value proxy group. HI zones were also closer to the urban center and had greater average subway-network density. Because the clusters were partly characterized using the response indicators and because demographic and infrastructure characteristics were not randomly assigned, these patterns should not be interpreted as evidence that population composition or transit infrastructure independently caused the larger estimated responses. Together, these profiles indicate that larger program-associated responses were concentrated in zones combining greater transit accessibility with particular user and travel profiles, without isolating demographic or infrastructure causality.

**Discussion**

This study provides large-scale real-world evidence that enrollment in a digital carbon-incentive program was associated with changes in observed travel behavior. The study integrates 4.82 billion trip records from 2.96 million observed platform users with a matched longitudinal panel of 632,376 CIP-MaaS participants and 1,264,753 non-enrolled platform users. The matched difference-in-differences analysis estimated a 20.3 percentage-point increase in participants' monthly low-carbon travel ratio, with the staggered DID robustness specification yielding a model-based average post-enrollment ATT of 19.5 percentage points. A secondary random-forest accounting analysis produced a model-implied 1.8% citywide decline in gasoline-car trips and an annualized gross carbon-accounting value of 94,353 tonnes $CO_2$ , equivalent in scale to 5.7% of certified reductions traded in Beijing in 2023. The behavioral estimates are interpreted as quasi-experimental program-associated changes, whereas the carbon estimates remain conditional on the specified model and accounting assumptions.

These findings contribute to research on MaaS, behavioral incentives, and demand-side transport decarbonization. Much of the existing evidence on MaaS and mobility incentives comes from pilot programs, surveys, or stated-preference data, which limits inference on observed behavior at scale. By contrast, this study uses passively recorded longitudinal mobility data from an operational digital platform. The CIP-MaaS program also differs from conventional MaaS pilots by linking platform-based rewards to verified carbon accounting. It therefore illustrates how digital mobility platforms may function not only as trip-planning tools, but also as operational channels for carbon-linked behavioral interventions.

The estimated responses varied across both individuals and urban contexts. Larger program-associated reductions in gasoline-car use and $CO_2$ emissions were spatially concentrated in zones with greater subway-network density and shorter distances to the urban center, and cluster-specific regressions showed positive associations between subway-network density and both response indicators. This co-location is consistent with the intuition that incentives are most actionable where low-carbon alternatives already exist. However, because the spatial classification partly reflects the observed response patterns themselves, we interpret this as spatial concentration rather than a causal test of transit-accessibility moderation.

From a behavioral perspective, the observed responses are consistent with several plausible mechanisms, including reward salience, immediate feedback, integration into trip planning, and the availability of convenient low-carbon alternatives. While the present data do not allow us to fully disentangle these mechanisms or estimate precise behavioral elasticities with respect to reward size, travel time, travel cost, or mode availability, the consistency of the observed patterns with established behavioral channels is consistent with the interpretation that incentives contributed to the observed response. Similarly, although we cannot completely separate mode substitution from potential changes in trip frequency, destination choice, platform engagement, or recording intensity, the sustained elevation in the low-carbon travel ratio during the observation window indicates a sustained increase in the low-carbon share of observed trips while incentives were active, even if the exact behavioral pathways remain only partially identified.

Several limitations should be considered. First, participation in CIP-MaaS was voluntary rather than randomly assigned. Although the matched difference-in-differences design, event-study diagnostics, placebo tests, staggered DID estimates using not-yet-treated future participants, Oster coefficient-stability analysis, and external validation using public-transport transaction records reduce several alternative explanations, they do not eliminate the possibility of unobserved selection. In particular, the absence of statistically significant pre-enrollment event-study coefficients provides no evidence of large differential pre-enrollment trends within the observed window but does not prove that the parallel-trends assumption held exactly.

Second, individual monthly total travel emissions were not observed for both participants and matched non-participants, so we could not estimate an aggregate emission difference-in-differences outcome. The carbon-accounting values therefore rely on model-classified reference modes, including actual trip duration and distance as model inputs, and should be interpreted as scenario-based accounting estimates rather than direct difference-in-differences estimates of total emissions. Third, the analysis is based on one large digital mobility ecosystem in Beijing, which may limit generalizability to other cities, platforms, and transport systems. Fourth, because incentives remained active during the observation period, the study cannot determine whether behavioral responses would persist after incentives are removed, although the observed responses suggest that incentives can support continued engagement while they remain active.

The findings have implications for the design of demand-side transport decarbonization policies. Market-financed, platform-based incentives may help support low-carbon travel engagement when they are embedded in everyday mobility services and aligned with accessible public transport or cycling options. However, such programs are unlikely to be stand-alone solutions for urban transport decarbonization. They should instead be viewed as complementary demand-side instruments alongside low-carbon infrastructure, reliable mobility services, vehicle electrification, and transparent carbon-accounting systems. Further research should report incentive costs more explicitly and examine behavioral mechanisms, equity implications, and post-incentive durability.

## Methods

### Datasets

Our analysis integrates several de-identified data sources from the Gaode digital mobility platform, including trip-level mobility records, CIP-MaaS program-registration and carbon-account records, reward-redemption and financial-flow records, an independent public-transport transaction dataset, and public geographic data. The trip-level behavioral analysis covers December 1, 2022 to December 30, 2023. Users in the trip dataset are platform users with at least one valid mobility record during this observation window. CIP-MaaS registrants are users recorded in the program registry between December 2019 and December 2023. Not all cumulative registrants had valid trip records during the behavioral observation window. For the matched analysis, participants were defined as observed platform users who enrolled in CIP-MaaS, whereas comparison users were observed platform users who had not yet enrolled during the corresponding matching and comparison period.

The main trip-level dataset contained de-identified user IDs, departure times, origin and destination coordinates, travel distances, travel durations, dates, and inferred travel modes. After removing records with missing key fields or locations outside the Beijing municipal boundary, the final trip dataset comprised 4,824,377,245 valid trip records.

We supplemented the mobility records with demographic and socioeconomic proxy information from a broader Gaode backend user table containing information on 14,708,023 platform accounts. The program registry identified 3,983,027 cumulative CIP-MaaS registrations between December 2019 and December 2023. These broader backend and registration counts describe the available source databases and should not be interpreted as the number of users represented in the final trip-level or matched analytical samples. Only records that could be linked to eligible users in the corresponding analytical dataset were used. The available variables included gender, age, and a categorical mobile-phone-value proxy derived from estimated mobile-phone value. This proxy was used as an observed covariate and does not measure reported individual or household income.

To construct the denominator of each user's monthly low-carbon travel ratio, we used auxiliary daily travel activity inferred from user-consented mobile phone location signals linked to Gaode mobility services. This dataset records anonymous user IDs, departure times, origins, destinations, and inferred travel modes. Location signals were collected only from users who had granted permission for location sharing while the relevant applications were in use or running in the background. Gaode processed these raw location signals using trajectory reconstruction and mode-inference algorithms to identify complete daily travel activity. By linking anonymous user IDs across the mobility and program datasets, we matched total observed trips with low-carbon trips recorded in the Gaode backend. The monthly low-carbon travel ratio was then defined as the share of low-carbon trips among all observed trips made by a user in a given month. This monthly aggregation was used for the individual-level DID analyses because it provides a stable denominator for repeated behavioral measurement.

We also used CIP-MaaS carbon-account records for registered participants. These records include de-identified user IDs, registration times, trip departure times, and credited carbon-emission reductions for eligible low-carbon trips. By matching user IDs and departure times, we identified the travel modes associated with credited reductions. These carbon-account records describe program-recorded credited reductions, but they do not by themselves distinguish program-induced mode shifts from low-carbon trips that participants might have taken without the incentive. For this reason, they were used for program accounting and carbon-reduction estimation in combination with the counterfactual mode-inference procedure described below.

To characterize treatment magnitude and the funding channel, we incorporated certified PCER transaction records and Gaode reward-redemption records. We linked each valid reward-redemption record to the corresponding participant's enrollment month and calculated event month as the difference, in calendar months, between the redemption month and enrollment month. Event month 0 denoted the calendar month of enrollment, event month 1 denoted the first calendar month after enrollment, and so forth. Reward exposure was summarized over event months 0–8 at the participant-month level. These financial data were used descriptively to quantify incentive scale and funding alignment rather than to estimate behavioral responses to reward size.

To evaluate the reliability of platform-based public-transport records and assess whether the estimated program response could plausibly be explained by changes in app usage or recording intensity, we used an independent reference dataset provided by the Beijing Public Transport Corporation. This dataset consists of QR-code and smartcard transaction records for public transport use. We conducted stratified random sampling by month and user group. For each month from December 2022 to October 2023, we extracted monthly low-carbon trip counts for sampled non-participants. For each enrollment cohort from April to October 2023, we applied the same procedure to participants before and after joining CIP-MaaS. We then compared trip-frequency distributions between Gaode backend records and public-transport transaction records using paired tests and equivalence tests. Because this validation dataset records public-transport transactions independently from the Gaode app backend, it is most informative for validating public-transport-related low-carbon trip recording rather than all travel modes equally.

To examine spatial heterogeneity, we incorporated geographic data for Beijing from OpenStreetMap (2023), including road, bus, and subway network information. Beijing was divided into 500 m × 500 m grid zones. Trip origins and destinations were assigned to these zones using their coordinates. Zone-level transport-network density was calculated from the length of road, bus, and subway networks within each zone. For the spatial response analysis, trips were assigned to both origin and destination zones and used to construct directed inter-zone mobility graphs; zone-level trip counts, credited reductions, and mode-specific travel-time and distance indicators were then summarized according to the corresponding origin and destination roles in the model specification. This zone-based aggregation was used to examine how observed program responses varied with local transport-infrastructure conditions.

All trip, carbon-account, and auxiliary location data were analyzed in de-identified form. Individual identifiers were anonymized before analysis, and exact trip origins and destinations were spatially aggregated to grid-zone centroids to reduce re-identification risk. The auxiliary location dataset used to reconstruct total travel exposure was derived from user-consented mobile phone location signals. All analyses were conducted within a secure data environment provided by the platform operator, and raw identifiable data were not exported to the research team. Data processing followed the platform's data-security and privacy-management protocols and complied with applicable data-governance requirements.

**Estimation of program-associated changes in individual low-carbon travel**

Because participation in CIP-MaaS was not randomly assigned, direct comparisons between participants and non-participants may be affected by selection bias. To improve comparability between the two groups, we used propensity score matching based on observed socio-demographic and travel-related characteristics[36–39]. Specifically, participants and non-participants were matched on gender, age, mobile-phone-value proxy, average travel time, and number of trips using nearest-neighbour matching with a 1:2 ratio and a caliper of 0.25 standard deviations of the propensity score. Matching was implemented separately by enrollment period because users joined CIP-MaaS at different times. Matching quality was assessed using chi-square tests and standardized comparisons of covariate balance across groups[38]. After matching, observations were combined across enrollment periods to construct a matched longitudinal panel. The resulting matched panel contained 632,376 enrolled CIP-MaaS participants and 1,264,753 non-enrolled platform users, corresponding to 1,897,129 unique users and 17,841,159 user-month observations. The matched participants represent a subset of cumulative CIP-MaaS registrants with eligible mobility records and the covariate information required for matching.

The main outcome variable was the monthly low-carbon travel ratio, defined as the share of low-carbon trips among all observed trips made by an individual in a given month. The unit of analysis was therefore the individual-month observation. We estimated program-associated changes using a time-varying DID framework with individual and month fixed effects. Standard errors were clustered at the individual level. The baseline model was specified as:

$$PST_{it} = \pi_i + \rho_t + \delta * DP_{it} + \epsilon_{it} \qquad (1)$$

where $PST_{it}$ denotes the monthly low-carbon travel ratio of individual $i$ in month $t$. $DP_{it}$ is an indicator equal to 1 if the individual had enrolled in the CIP-MaaS program by month $t$ and 0 otherwise. $\pi_i$ is the individual-related fixed effect, which controls all the individual-related factors that do not change with time; $\rho_t$ is the time-fixed effect, which controls all the time-related factors that are not individual specific. $\epsilon_{it}$ is the error term; $\delta$ captures the program-associated change in the monthly low-carbon travel ratio after enrollment, conditional on the fixed effects. Accordingly, estimates based on this outcome should be interpreted as changes in the low-carbon share of observed trips, rather than as a complete decomposition of changes in mode choice, trip generation, or destination choice.

To examine pre-enrollment dynamics relevant to the parallel-trends assumption, we estimated an event-study specification with relative-time indicators centered on each participant's enrollment month:

$$PST_{it} = \pi_i + \rho_t + \sum_{k \geq -4, k \neq -1}^{k=8} \beta_k DP_{it,k} + \epsilon_{it} \qquad (2)$$

where $DP_{it,k}$ represents a series of indicator variables that reflect whether an individual participated in the incentive program during various periods. Specifically, $DP_{it,k}$ equals zero for all periods if individual $i$ never participated in this incentive program. Conversely, if individual $i$ joined the program in a particular period $t$, with $ST$ representing the time at which an individual started participating in this program, then $DP_{it,k} = 1$ for any period $k$ that occurs after or during $t - ST$ and zero for any period prior to that time. By considering the timing of the start of participation, we could test whether there are any differences in trends prior to implementing this program among individuals in the treatment and control groups. Pre-enrollment coefficients were interpreted as diagnostic evidence on differential trends; failure to detect statistically significant pre-enrollment coefficients was not treated as proof that the parallel-trends assumption held exactly.

To examine heterogeneous responses across socio-demographic and travel-behavior groups, we estimated subgroup-specific DID models. Individuals in both the treatment and control groups were categorized according to gender, age group, mobile-phone-value proxy, monthly trip frequency, and average travel time per trip. The interaction model was specified as:

$$PST_{it} = \pi_i + \rho_t + \beta_0 DP_{it} + \sum_c \beta_c DP_{it} V_c + \epsilon_{it} \quad (3)$$

where $\beta_0$ is of primary interest, quantifying the average effect of the intervention. $V_c$ represents the covariate variables, allowing the model to account for variations in the incentive effect based on different individual sub-characteristics. The interaction terms $\beta_c$ captures how the treatment effects vary across different subgroups. These models were used to characterize heterogeneity in the estimated response rather than to identify individual-level treatment effects.

As an additional robustness check, we conducted placebo analyses by assigning hypothetical enrollment dates one or two months earlier than the observed enrollment month and re-estimating the DID specification. The absence of significant placebo effects supports the interpretation that the main results are not driven by spurious timing or common external shocks. To further assess the sensitivity of the results to voluntary enrollment, we estimated a staggered DID specification using not-yet-treated future participants as the comparison group[40,41]. In this analysis, users who had already enrolled in CIP-MaaS were compared with users who would enroll later but had not yet joined in the same calendar month. Treatment timing was defined as the first month of CIP-MaaS enrollment, and the outcome was the individual monthly low-carbon travel ratio, expressed in percentage points. Dynamic estimates were calculated relative to the month immediately before enrollment. Details of the estimator, sample construction, and event-study results are reported in Note S4.

We further used Oster coefficient-stability analysis to assess how strong selection on unobserved variables would need to be to explain away the DID estimate[42]. This analysis compared an uncontrolled two-way fixed-effects DID specification with a controlled specification that additionally included observed socio-demographic characteristics and pre-enrollment travel characteristics interacted with month indicators. Details are reported in Note S4.

**Scenario-based carbon accounting using model-classified reference modes**

We conducted a secondary trip-level classification and accounting analysis to characterize the potential carbon implications of observed post-enrollment travel patterns. Monthly carbon-account records report credited reductions from eligible low-carbon trips but do not identify the travel mode that would have been selected without program enrollment. We therefore trained a random-forest multiclass classifier using participants' pre-enrollment trips and applied it to post-enrollment trips to generate a model-classified reference mode[43].

For each post-enrollment trip, we compared the observed travel mode with the model-classified reference mode. A trip contributed to the carbon-accounting total when the classified reference mode was gasoline car and the observed mode was bus, subway, or cycling.

The corresponding emission difference was calculated using the mode-specific accounting procedure described in Note S2. The resulting values are conditional on observed trips, model inputs, classification rules, and emission factors. They do not capture possible program-associated changes in trip generation, destination choice, departure timing, or trip chaining.

The random-forest model used departure time, origin, destination, realized trip duration, realized trip distance, workday status, and travel date as input features. The output was a model-classified reference mode—gasoline car, bus, subway, or cycling—for each trip. The model was trained using participants' pre-enrollment travel records only. Trip records were randomly divided into training and testing samples, with 80% used for model training and 20% reserved for predictive evaluation; five-fold cross-validation was also conducted. Because the split was implemented at the trip level, the training and testing samples could contain trips from the same users or similar origin–destination pairs. In addition, realized duration and distance may contain information associated with the observed mode. The reported performance measures should therefore be interpreted as predictive classification diagnostics within the observed data rather than validation of untreated causal counterfactuals. Precision, recall, and F1-score were calculated separately for each mode class using a one-versus-rest representation[44]:

$$Ac = \frac{TP+TN}{TP+FN+FP+TN} \tag{4}$$

$$Re = \frac{TP}{TP+FN} \tag{5}$$

$$Pr = \frac{TP}{TP+FP} \tag{6}$$

$$F_1 = \frac{2Pr \times Re}{Pr+Re} \tag{7}$$

where $AC$ is the accuracy, which is the proportion of correctly classified samples out of the total number of samples. $Re$ is the recall, which measures the model's ability to correctly identify positive samples (high values indicate satisfactory detection of positive cases). $Pr$ is the precision, reflecting the model's accuracy in predicting positive samples. High precision indicates that the model accurately identifies positive samples. $F_1$ is the harmonic mean of precision and recall, providing a balanced measure of model performance. $TP$, $TN$, $FP$, and $FN$ are the numbers of true positives, true negatives, false positives, and false negatives, respectively.

Next, we applied the trained random forest model to infer the preferred travel modes for each trip taken by individuals during the period from December 1, 2022, to December 30, 2023, following their participation in the CIP-MaaS program. We calculated three indicators to summarize inferred mode-shift and carbon-accounting outcomes: the mode-shift ratio, the reduction in gasoline-car travel ratio, and cumulative scenario-based carbon-accounting values (Note S5). These indicators are formulated as follows:

$$MS_{tij} = \frac{PA_{tij}}{P_{ti}} \tag{8}$$

$$RCT_t = 1 - \frac{PAC_t}{PC_t} \tag{9}$$

$$CER_t = \sum_m \sum_k ER_{car \to m}(t,k) \tag{10}$$

where $MS_{tij}$ denotes the share of trips observed in mode $j$ among trips assigned reference mode $i$ on day $t$. $PA_{tij}$ is the number of trips observed in mode $j$ that were assigned reference mode $i$ by the classifier, and $P_{ti}$ is the total number of trips assigned reference mode $i$. $RCT_t$ denotes the model-implied reduction in gasoline-car travel under the specified accounting procedure. $PAC_t$ is the number of trips observed as gasoline-car travel among trips assigned a gasoline-car reference mode, and $PC_t$ is the total number of trips assigned a gasoline-car reference mode. $CER_t$ denotes the cumulative scenario-based carbon-accounting value for trips assigned a gasoline-car reference mode but observed in a low-carbon mode. $CER_{car \to m}(t,k)$ denotes the accounting value for the $k$-th such trip observed in mode m, where m includes bus, subway, and bicycle (Note S2).

**Analysis of urban characteristics associated with program responses**

To characterize spatial heterogeneity in program-associated responses, Beijing was divided into 500 m × 500 m grid zones. Each trip was assigned to both an origin zone and a destination zone according to its origin and destination coordinates. We constructed daily directed mobility graphs in which zones were represented as nodes and inter-zone trips were represented as edges. Edges were defined separately by travel mode, including gasoline car, bus, subway, and bicycle. Node features summarized daily modal volumes, average travel times, and average travel distances associated with each zone.

We used a dual-channel graph convolutional network to learn zone embeddings from these daily multimodal mobility graphs[45]. The two channels encoded zones separately according to their roles as trip origins and trip destinations, allowing the model to capture directional mobility patterns. The model output included zone-level indicators of estimated shifts from gasoline cars to low-carbon modes, reductions in gasoline-car travel, and estimated $CO_2$ reductions. The data were split into training and test sets, and early stopping and dropout were used to reduce overfitting. Model performance was evaluated using mean squared error, root mean squared error, and mean absolute error.

The learned zone embeddings were projected using UMAP for visualization[47] and grouped using hierarchical clustering[46]. The Calinski – Harabasz index[48] and silhouette coefficient[49] supported a four-cluster solution among the candidate specifications. For descriptive interpretation, the clusters were labeled negligible response, low response, moderate response, and high response according to the empirical distributions of gasoline-car travel reduction, scenario-based $CO_2$ accounting values, and mode-shift indicators. Because these response indicators contributed to the interpretation and labeling of the clusters, subsequent differences in the same indicators across clusters were treated as descriptive features of the grouping procedure rather than independent evidence of heterogeneous causal effects.

To examine urban characteristics correlated with the spatial patterns, we fitted separate ordinary least-squares regressions within each cluster. The dependent variables were zone-level gasoline-car travel differences and scenario-based $CO_2$ accounting values. Explanatory variables included road, bus, and subway network densities and average travel times by gasoline car and public transport.

**Ethics declarations**

Human Ethics and Consent to Participate: This study used de-identified, retrospective mobility and platform-usage data provided by Gaode (AutoNavi) under its existing user agreement and privacy policy. The research team did not directly recruit, contact, or intervene with any human participants at any stage of the study. Consent for the collection and research use of anonymized platform data was obtained by Gaode (AutoNavi) from users at the time of app registration and use, in accordance with its terms of service and privacy policy, and in compliance with the Data Security Law of the People's Republic of China and Alibaba's Data Security Management Standards. All data were anonymized prior to analysis, exact trip origins and destinations were spatially aggregated to reduce re-identification risk, and the research team had no access to identifiable personal information at any stage.

In accordance with Article 32 of the Measures for Ethical Review of Life Science and Medical Research Involving Human Beings (2023), which exempts from formal ethical review research that uses legally obtained, anonymized data or non-interventional observational data not involving sensitive personal information, this study was exempted from full ethical committee review. This exemption was confirmed by the School of Transportation Science and Engineering, Beihang University, on 3 Jan2026.

**Data availability**

The multimodal travel data, carbon-account data, and reward-redemption records supporting this study were provided by the Gaode app service provider, a research collaborator, and were analyzed under applicable data-security, privacy, and disclosure protocols. These data comply with the Data Security Law of the People's Republic of China, Alibaba's Data Security Management Standards, and its Data Disclosure Protocol. Due to strict data privacy and disclosure regulations, these datasets are not publicly available. However, aggregated data and visualizations of key results, enabling the reproduction of findings, are publicly accessible at: https://github.com/none1113350/carbon_incentive_Beijing.git.

Beijing's transportation statistics were sourced from the 2024 Beijing Annual Transportation Development Report (https://www.bjtrc.org.cn/List/index/cid/7.html), which is publicly available. Data from Beijing's carbon trading market were obtained from the Beijing Carbon Emission Trading Electronic Platform (https://www.bjets.com.cn/), which is also publicly accessible. All data were used in compliance with the service providers' specified terms of use.

**References**

1. Solecki, W., Roberts, D., and Seto, K.C. (2024). Strategies to improve the impact of the IPCC Special Report on Climate Change and Cities. Nat. Clim. Chang. *14*, 685–691. https://doi.org/10.1038/s41558-024-02060-9.
2. Sovacool, B.K., Upham, P., Martiskainen, M., Jenkins, K.E.H., Torres Contreras, G.A., and Simcock, N. (2023). Policy prescriptions to address energy and transport poverty in the United Kingdom. Nat. Energy *8*, 273–283. https://doi.org/10.1038/s41560-023-01196-w.
3. Xiao, C., Sluijs, E. van, Ogilvie, D., Patterson, R., and Panter, J. (2022). Shifting towards healthier transport: carrots or sticks? Systematic review and meta-analysis of population-level interventions. Lancet Planet. Heal. *6*, e858–e869. https://doi.org/10.1016/S2542-5196(22)00220-0.
4. Fu, X., Cheng, J., Peng, L., Zhou, M., Tong, D., and Mauzerall, D.L. (2024). Co-benefits of transport demand reductions from compact urban development in Chinese cities. Nat. Sustain. *7*, 294–304. https://doi.org/10.1038/s41893-024-01271-4.
5. Javaid, A., Creutzig, F., and Bamberg, S. (2020). Determinants of low-carbon transport mode adoption: systematic review of reviews. Environ. Res. Lett. *15*, 103002. https://doi.org/10.1088/1748-9326/aba032.
6. Winkler, L., Pearce, D., Nelson, J., and Babacan, O. (2023). The effect of sustainable mobility transition policies on cumulative urban transport emissions and energy demand. Nat. Commun.

*14*, 1–14. https://doi.org/10.1038/s41467-023-37728-x.
7. Liotta, C., Viguié, V., and Creutzig, F. (2023). Environmental and welfare gains via urban transport policy portfolios across 120 cities. Nat. Sustain. 2023 69 *6*, 1067–1076. https://doi.org/10.1038/s41893-023-01138-0.
8. Jittrapirom, P., Caiati, V., Feneri, A.M., Ebrahimigharehbaghi, S., Alonso-González, M.J., and Narayan, J. (2017). Mobility as a service: A critical review of definitions, assessments of schemes, and key challenges. Urban Plan. *2*, 13–25. https://doi.org/10.17645/up.v2i2.931.
9. Ramaswami, A., Tong, K., Canadell, J.G., Jackson, R.B., Stokes, E. (Kellie), Dhakal, S., Finch, M., Jittrapirom, P., Singh, N., Yamagata, Y., et al. (2021). Carbon analytics for net-zero emissions sustainable cities. Nat. Sustain. *4*, 460–463. https://doi.org/10.1038/s41893-021-00715-5.
10. Tsouros, I., Tsirimpa, A., Pagoni, I., and Polydoropoulou, A. (2021). MaaS users: Who they are and how much they are willing-to-pay. Transp. Res. Part A Policy Pract. *148*, 470–480. https://doi.org/10.1016/j.tra.2021.04.016.
11. Technology, I.A.C. (2024). Global Mobility As A Service Market – Industry Trends and Forecast to 2031.
12. Reyes Madrigal, L.M., Nicolaï, I., and Puchinger, J. (2023). Pedestrian mobility in Mobility as a Service (MaaS): sustainable value potential and policy implications in the Paris region case. Eur. Transp. Res. Rev. *15*, 1–20. https://doi.org/10.1186/s12544-023-00592-3.
13. Butler, L., Yigitcanlar, T., and Paz, A. (2021). Barriers and risks of Mobility-as-a-Service (MaaS) adoption in cities: A systematic review of the literature. Cities *109*, 103036. https://doi.org/10.1016/j.cities.2020.103036.
14. Intergovernmental Panel on Climate Change AR6 Synthesis Report: Climate Change 2023. https://www.ipcc.ch/report/sixth-assessment-report-cycle/.
15. Kriswardhana, W., and Esztergár-Kiss, D. (2023). Exploring the aspects of MaaS adoption based on college students' preferences. Transp. Policy *136*, 113–125. https://doi.org/10.1016/j.tranpol.2023.03.018.
16. Frank, L., Klopfer, A., and Walther, G. (2024). Designing corporate mobility as a service – Decision support and perspectives. Transp. Res. Part A Policy Pract. *182*, 104011. https://doi.org/10.1016/j.tra.2024.104011.
17. Morfeldt, J., and Johansson, D.J.A. (2022). Impacts of shared mobility on vehicle lifetimes and on the carbon footprint of electric vehicles. Nat. Commun. *13*, 1–11. https://doi.org/10.1038/s41467-022-33666-2.
18. Diao, M., Kong, H., and Zhao, J. (2021). Impacts of transportation network companies on urban mobility. Nat. Sustain. *4*, 494–500. https://doi.org/10.1038/s41893-020-00678-z.
19. Nikitas, A., Cotet, C., Vitel, A.E., Nikitas, N., and Prato, C. (2024). Transport stakeholders' perceptions of Mobility-as-a-Service: A Q-study of cultural shift proponents, policy advocates and technology supporters. Transp. Res. Part A Policy Pract. *181*, 103964. https://doi.org/10.1016/j.tra.2024.103964.
20. Ydersbond, I.M., Auvinen, H., Tuominen, A., Fearnley, N., and Aarhaug, J. (2020). Nordic experiences with smart mobility: Emerging services and regulatory frameworks. Transp. Res. Procedia *49*, 130–144. https://doi.org/10.1016/j.trpro.2020.09.012.
21. Nisa, C.F., Bélanger, J.J., Schumpe, B.M., and Faller, D.G. (2019). Meta-analysis of randomised controlled trials testing behavioural interventions to promote household action on

climate change. Nat. Commun. *10*, 1–13. https://doi.org/10.1038/s41467-019-12457-2.

22. Hoerler, R., Stünzi, A., Patt, A., and Del Duce, A. (2020). What are the factors and needs promoting mobility-as-a-service? Findings from the Swiss Household Energy Demand Survey (SHEDS). Eur. Transp. Res. Rev. *12*, 1–16. https://doi.org/10.1186/s12544-020-00412-y.
23. ITS International (2018). Whim launch in Birmingham: new day dawning. https://www.itsinternational.com/its17/feature/whim-launch-birmingham-new-day-dawning.
24. Caiati, V., Rasouli, S., and Timmermans, H. (2020). Bundling, pricing schemes and extra features preferences for mobility as a service: Sequential portfolio choice experiment. Transp. Res. Part A Policy Pract. *131*, 123–148. https://doi.org/10.1016/j.tra.2019.09.029.
25. Zhang, R., and Hanaoka, T. (2022). Cross-cutting scenarios and strategies for designing decarbonization pathways in the transport sector toward carbon neutrality. Nat. Commun. *13*. https://doi.org/10.1038/s41467-022-31354-9.
26. Ho, C.Q., Mulley, C., and Hensher, D.A. (2020). Public preferences for mobility as a service: Insights from stated preference surveys. Transp. Res. Part A Policy Pract. *131*, 70–90. https://doi.org/10.1016/j.tra.2019.09.031.
27. Tan-Soo, J.S., Qin, P., Quan, Y., Li, J., and Wang, X. (2024). Using cost–benefit analyses to identify key opportunities in demand-side mitigation. Nat. Clim. Chang., 1–7. https://doi.org/10.1038/s41558-024-02146-4.
28. Koch, N., Naumann, L., Pretis, F., Ritter, N., and Schwarz, M. (2022). Attributing agnostically detected large reductions in road CO2 emissions to policy mixes. Nat. Energy *7*, 844–853. https://doi.org/10.1038/s41560-022-01095-6.
29. Plötz, P., Axsen, J., Funke, S.A., and Gnann, T. (2019). Designing car bans for sustainable transportation. Nat. Sustain. *2*, 534–536. https://doi.org/10.1038/s41893-019-0328-9.
30. Gao, K., Jia, R., Liao, Y., Liu, Y., Najafi, A., and Attard, M. (2024). Big-data-driven approach and scalable analysis on environmental sustainability of shared micromobility from trip to city level analysis. Sustain. Cities Soc. *115*, 105803. https://doi.org/10.1016/j.scs.2024.105803.
31. Beijing Daily The city encourages citizens to go green in a carbon-inclusive way. People's Gov. Beijing Munic. https://www.beijing.gov.cn/renwen/sy/whkb/202009/t20200911_2057908.html#.
32. Su Song, Miaoqing Zhong, D.T. (2023). How Mobility-as-a-Service Platforms Can Pilot Greener Travel Behaviors. World Resour. Inst. https://thecityfix.com/blog/how-mobility-as-a-service-can-encourage-greener-travel/.
33. Beijing Municipal Commision of Transport (2020). Beijing's MaaS platform launched the "MaaS Travel, Green City" initiative, which is the first in China to encourage citizens to participate in green travel in all ways in a carbon-inclusive way. Beijing Munic. Commision Transp. https://jtw.beijing.gov.cn/xxgk/dtxx/202009/t20200908_1999634.html.
34. Beijing Municipal Commision of Transport (2020). The incentive effect of green travel has begun to appear, and a total of 2.45 million people have been served. Beijing Munic. Commision Transp. https://jtw.beijing.gov.cn/xxgk/xwfbh/202011/t20201103_2127960.html.
35. China, B.G.E. Beijing Carbon Emissions Trading Platform. https://www.bjets.com.cn/.
36. Kato, H. (2024). Daily walking time effects of the opening of a multifunctional facility "ONIKURU" using propensity score matching and GPS tracking techniques. Sci. Reports 2024 141 *14*, 31047. https://doi.org/10.1038/s41598-024-82232-x.
37. Yılmaz, Ö., Son, Y., Shang, G., and Arslan, H.A. (2024). Causal inference under selection on

observables in operations management research: Matching methods and synthetic controls. J. Oper. Manag. *70*, 831–859. https://doi.org/10.1002/JOOM.1318.

38. Austin, P.C. (2011). An Introduction to Propensity Score Methods for Reducing the Effects of Confounding in Observational Studies. Multivariate Behav. Res. *46*, 399–424. https://doi.org/10.1080/00273171.2011.568786.
39. Rosenbaum, P.R., and Rubin, D.B. (1983). The central role of the propensity score in observational studies for causal effects. Biometrika *70*, 41–55. https://doi.org/10.1093/BIOMET/70.1.41.
40. Roth, J., Sant'Anna, P.H.C., Bilinski, A., and Poe, J. (2023). What's trending in difference-in-differences? A synthesis of the recent econometrics literature. J. Econom. *235*, 2218–2244. https://doi.org/10.1016/J.JECONOM.2023.03.008.
41. Callaway, B., and Sant'Anna, P.H.C. (2021). Difference-in-Differences with multiple time periods. J. Econom. *225*, 200–230. https://doi.org/10.1016/J.JECONOM.2020.12.001.
42. Oster, E. (2019). Unobservable Selection and Coefficient Stability: Theory and Evidence. J. Bus. Econ. Stat. *37*, 187–204. https://doi.org/10.1080/07350015.2016.1227711.
43. Breiman, L. Random forests. Machine Learning 45, 5–32 (2001).
44. Sokolova, M., and Lapalme, G. (2009). A systematic analysis of performance measures for classification tasks. Inf. Process. Manag. *45*, 427–437. https://doi.org/10.1016/j.ipm.2009.03.002.
45. Kipf, T.N., and Welling, M. (2017). Semi-Supervised Classification with Graph Convolutional Networks.
46. Ward, J.H. (1963). Hierarchical Grouping to Optimize an Objective Function. J. Am. Stat. Assoc. *58*, 236–244. https://doi.org/10.1080/01621459.1963.10500845.
47. McInnes, L., Healy, J., and Melville, J. (2018). UMAP: Uniform Manifold Approximation and Projection for Dimension Reduction.
48. Calinski, T., and Harabasz, J. (1974). A dendrite method foe cluster analysis. Commun. Stat. *3*, 1–27. https://doi.org/10.1080/03610927408827101.
49. Rousseeuw, P.J. (1987). Silhouettes: A graphical aid to the interpretation and validation of cluster analysis. J. Comput. Appl. Math. *20*, 53–65. https://doi.org/10.1016/0377-0427(87)90125-7.

**Acknowledgements**

This work was supported by the Beijing Natural Science Foundation (No. JQ24051), and the European Research Council under the European Union's Horizon 2020 research and innovation programme (grant agreement No. 804469).

**Author contributions**

B.L. designed the study, analysed the data and wrote the first draft of the paper. Y.L., S.Y., and O.C. designed the study, supervised all parts of the analysis, analysed the data and edited the paper. Z.N.D. carried out data preparation and provided the data analysis platform. K.S.N., Y.W., Y.L.L, and Z.R.N. edited the paper. X.L.M. designed the study, analysed the data, supervised all parts of the analysis and directed the study. All authors read and commented on the final manuscript.

**Competing interests**

The authors declare no competing interests.

**Declaration of generative AI and AI-assisted technologies in the writing process**

During the preparation of this work, the authors used ChatGPT to assist with language editing. The authors reviewed and edited all generated content and take full responsibility for the content of the publication.